\documentclass[11pt]{article}

\usepackage[a4paper, margin=2.5cm]{geometry}
\usepackage{amsmath, amssymb, amsthm}
\usepackage{graphicx}
\usepackage[english]{babel}
\usepackage[utf8]{inputenc}
\usepackage{xcolor}
\usepackage{tabularx}
\usepackage{siunitx}
\usepackage{multirow}

\usepackage{hyperref}
\hypersetup{
	colorlinks=true,
	linkcolor=blue,
	citecolor=blue,
	urlcolor=blue
}

\begin{document}
	
\interfootnotelinepenalty=10000
	
\title{Guided elastic waves for soft elastomer characterization: an alternative to conventional rheometry}
\author{S. Croquette, P. Chantelot, D.A. Kiefer,  C.Prada, F.Lemoult}
\maketitle
	
\begin{abstract}
\noindent
Elastic wave propagation is intrinsically sensitive to the mechanical properties of the medium through which it travels. In soft elastomers, this makes guided elastic waves natural probes of viscoelastic and acoustoelastic behavior over a broad frequency range. In this work, we introduce a wave-based mechanical characterization method in which a thin elastomer strip acts as a waveguide supporting multiple in-plane guided modes.  
By combining stroboscopic measurements of monochromatic wave fields with a theoretical framework that couples frequency-dependent viscoelasticity and elongation-dependent acoustoelasticity, we extract complex-valued dispersion relations for guided modes under controlled static elongation. A dedicated numerical implementation allows these experimental dispersion curves to be quantitatively matched to theory, enabling identification of the material's rheological and hyperelastic parameters. 
Applied to several commercial silicone elastomers, the method yields mechanical parameters that are consistent with conventional plate-plate rheometry, while extending the accessible frequency range beyond that of conventional techniques. By exploiting the richness of guided-wave dispersion and the sensitivity of waves to both frequency and pre-stress, this approach provides a unified, broadband, and experimentally simple route to the mechanical characterization of soft elastomers.
\end{abstract}

\section{Introduction}

The mechanical characterization of soft materials is traditionally performed using quasi-static mechanical tests, such as uniaxial tension and compression, or dynamic tests, like shear rheometry. 
In these approaches, elastic and viscoelastic moduli are determined by simultaneously measuring the applied force and the resulting deformation, requiring accurate force sensing, precise boundary conditions, and careful calibration~\cite{treloar_2005,destrade2017methodical,meissner1985rheometry,osswald2014polymer}. 
These constraints can become limiting when probing soft or highly dissipative materials, small sample volumes, or non-standard geometries.

An alternative route to mechanical characterization relies on wave propagation. 
In a homogeneous elastic medium, the velocity and attenuation of elastic waves are directly governed by the material’s mechanical moduli and density~\cite{auld1973acoustic,landau1986,royer1999elastic}. 
As a consequence, mechanical parameters can be inferred from purely kinematic measurements of the wave field, without the need for direct force measurements. 
This intrinsic advantage makes wave-based methods particularly appealing for soft materials, where force measurements are often challenging or intrusive.
When waves are confined by a finite geometry, such as for plates, rods, or strips, the elastic field no longer propagates as a plane wave in a bulk but decomposes onto a set of guided modes~\cite{royer1999elastic}. 
Each mode is characterized by its own dispersion relation, spatial structure, and attenuation.
It interacts differently with the material’s mechanical properties and the guiding dimensions. 
From an experimental standpoint, this modal richness substantially increases the number of data points accessible from a single measurement configuration. 
Guided elastic waves therefore provide a powerful framework for material characterization, as multiple propagation modes can be simultaneously exploited over a broad frequency range to constrain the mechanical parameters of the medium~\cite{delory_2022}.

In this article, we develop a wave-based approach to characterize the frequency-dependent viscoelastic properties of soft elastomers using guided elastic waves. 
The material is shaped into a thin strip and directly used as a waveguide, enabling the extraction of mechanical parameters from the dispersion and attenuation of multiple guided modes. 
Particular attention is paid to the influence of static elongation, as the dynamic response of elastomers is strongly affected by pre-stress~\cite{ogden_1997,destrade_2007}, and to its coupling with viscoelasticity~\cite{delory_2023a,berjamin2025}.

The mechanical behavior of the material is described using viscoelastic constitutive models grounded in polymer physics and rheological modelling~\cite{meissner1985rheometry,osswald2014polymer,smit1970rheological,bagley1983fractional,mainardi2022fractional}. 
The corresponding model parameters are retrieved through a dedicated inversion procedure, allowing the material response to be quantified over a frequency range extending from 2~Hz to 300~Hz, with straightforward extension to higher frequencies.
The proposed framework relies on an incremental non-linear elasticity formulation combined with a semi-analytical computation of guided-wave dispersion relations~\cite{kiefer_elastodynamic_2022,kiefer_2024_10641373}. 
This approach fully accounts for the guiding geometry, the frequency-dependent viscoelastic properties of the elastomer, and their evolution under uniaxial pre-stress~\cite{destrade_2007,delory_2023a,delory_2023b,berjamin2025}.

Experimentally, guided elastic waves are generated in thin elastomeric strips molded from the target materials. 
This geometry allows the excitation and optical visualization of in-plane monochromatic guided shear waves using a conventional camera. 
Measurements are performed over a range of excitation frequencies and static elongations. The resulting wave fields are processed to extract dispersion relations, which are then compared to theoretical predictions to infer the mechanical parameters of the elastomer.

The method is validated on three representative elastomers and shows good agreement with standard rheological measurements. 
By combining broadband wave-based probing with a minimal experimental setup and explicit treatment of pre-stress effects, this approach offers a complementary and cost-effective alternative to conventional rheometry for the mechanical characterization of soft materials.

\section{Experiment}
We investigate three commercially available silicone rubbers from Smooth-On: Ecoflex~OO-30, Ecoflex~OO-50, and SortaClear~12. Ecoflex~OO-30 and Ecoflex~OO-50 have similar mechanical properties, providing a test of the method’s precision and reproducibility, while SortaClear~12 exhibits significantly different behavior, allowing us to evaluate the method’s robustness. %sensitivity to material variations.
For each elastomer, equal amounts of monomer and cross-linking agent are thoroughly mixed, poured into a horizontal rectangular mold, and left to cure at room temperature. After several hours, a translucent strip is obtained with dimensions $L = \qty{50}{\centi\meter}$, $b = \qty{5}{\centi\meter}$ and $h = \qty{2}{\milli\meter}$. To enable optical tracking of in-plane displacements, a fine black pigment is incorporated: the first layer of elastomer is poured, the pigment sprinkled on top, and a second layer is added to embed the markers within the material.

The objective of the experiments is to capture monochromatic guided waves propagating in these strips and to extract the corresponding dispersion relations, which serve as the basis for inferring the mechanical properties of the elastomers. The next section describes the experimental setup used to generate, control, and record these wave fields.

\subsection{Experimental setup}
\begin{figure}[t]
    \centering
    \includegraphics[width=8.7cm]{figure1.png}
    \caption{\textbf{Experimental setup ---} A strip is attached vertically to a motorized stage that stretches it along the direction $\boldsymbol{e}_1$. A point-like source connected to a shaker generates in-plane motion of the strip at a $\qty{45}{\degree}$ angle. A full-frame camera captures the resulting displacement of the strip.}
    \label{fig:setup}
\end{figure}

The experimental setup is illustrated in Figure~\ref{fig:setup}. 
Guided waves are generated by a point-like source oscillating at frequencies $f$ ranging from $\qty{2}{\hertz}$ to $\qty{300}{\hertz}$.
The source is polarized in the $(x_1,x_3)$ plane and rotated $\qty{45}{\degree}$ around the $x_2$ axis to simultaneously excite longitudinal and in-plane transverse modes.
In-plane displacements are captured using a full-frame camera (acA4112-20um, Basler) placed $\qty{2.5}{\meter}$ away from the strip. 
To overcome the camera's framerate limitation, stroboscopic imaging is used, so that, for each excitation frequency, we record a video consisting of sixty frames regularly spaced within a pseudo-period, similarly as in ref~\cite{lanoy_2020,delory_2022}. 
In this configuration, the frequency range is effectively limited only by the shaker.

To investigate the effect of pre-stress, the strip is clamped at its bottom edge to a motorized stage capable of imposing a uniaxial static elongation $\lambda_1$ along the longitudinal direction $x_1$. 
Assuming quasi-incompressibility, the transverse stretch ratios are linked to the longitudinal stretch ratio via $\lambda_2 = \lambda_3 = \lambda_1^{-0.5}$, and the applied elongation is only quantified by the scalar $\lambda_1$ which remains in the range $[1, 1.8]$.

This setup allows for optical tracking of the in-plane displacement of the strip under controlled excitation and pre-stretch, providing the data necessary to extract multiple guided-mode dispersion relations and subsequently infer the elastomer’s mechanical parameters.

\subsection{Dispersion curves extraction}
\label{section:postprocessing}
The main steps of the dispersion relation extraction procedure are summarized below; further details can be found in references~\cite{delory_2022,delory_2023b}.
\begin{figure}[t]
    \centering
    \includegraphics[width=\linewidth]{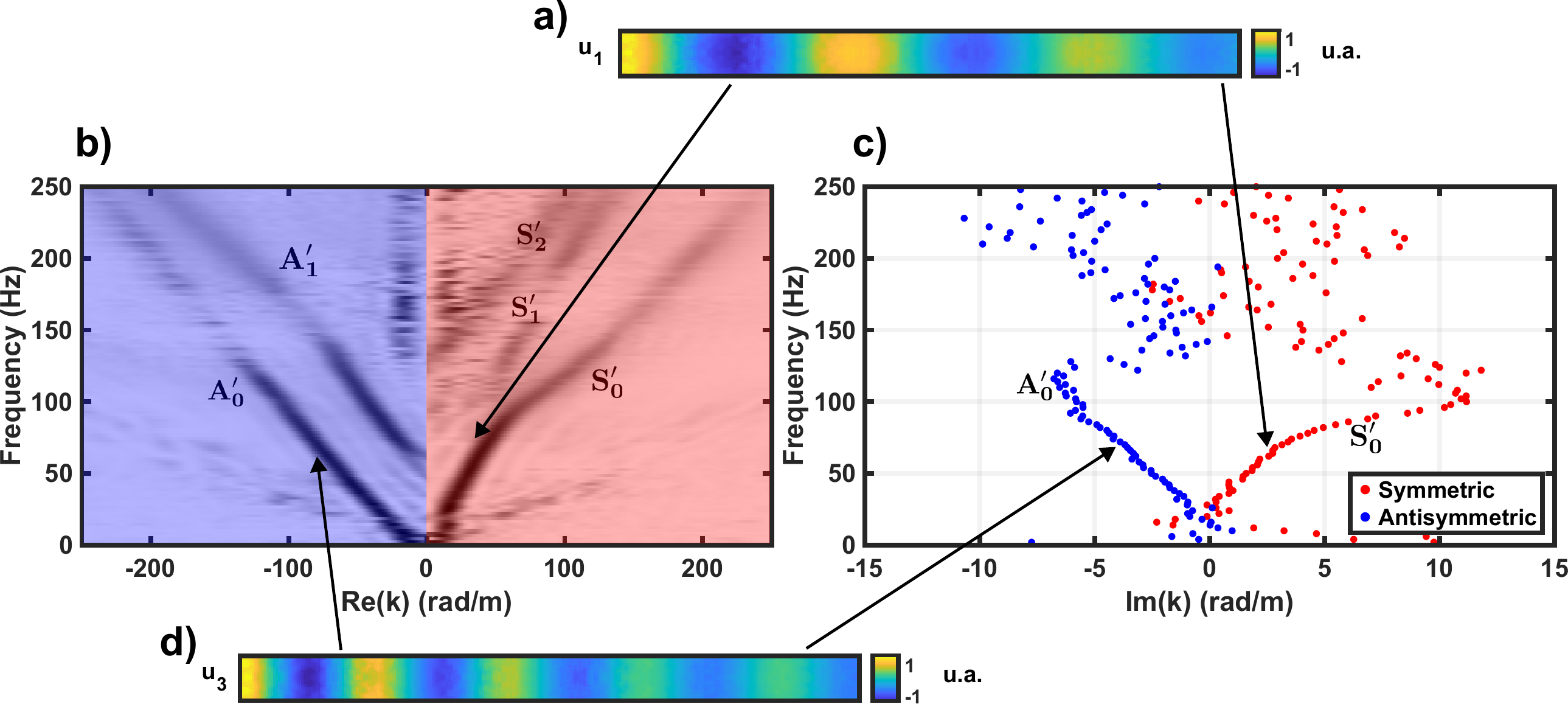}
    \caption{\textbf{Dispersion curves extraction ---} a,d) Relevant real displacement maps of the symmetric and antisymmetric modes at 70~Hz. b) Real part of the dispersion diagrams as intensity maps, extracted from the monochromatic displacement maps in the undeformed configuration ($\lambda_1 = 1$) for Ecoflex~OO-30. For clarity, symmetric and antisymmetric wavenumbers are plotted with opposite signs, although the corresponding waves propagate in the same physical direction. c) Imaginary counterparts, extracted using attenuation distances on the displacement maps.}
    \label{fig:displacement}
\end{figure}
For each imposed stretch ratio $\lambda_1$, each monochromatic excitation frequency $f = \omega / 2\pi$, and each recorded frame $n \in [1,N]$ with $N = 60$, the in-plane displacement field is obtained by Digital Image Correlation (DIC). 
The procedure outputs the full in-plane displacement maps $[u_1^{(n)}(x_1,x_3), u_3^{(n)}(x_1,x_3)]$ with sub-pixel resolution, where $(x_1, x_3)$ are the coordinates of a material point in the reference image, and $n$ is the frame number.
The complex monochromatic displacement field is then obtained by a discrete Fourier transform over time
\begin{equation}
u_i(x_1,x_3,\omega) = \frac{1}{N} \sum_{n=1}^{N} u_i^{(n)}(x_1,x_3)
e^{2\mathrm{i}\pi (n-1)/N},
\end{equation}
where $i \in {1,3}$ denotes the two in-plane displacement components.

The resulting displacement field over the full strip is decomposed into symmetric and antisymmetric components with respect to the mid-plane of the strip. An example of this decomposition is shown in Fig.~\ref{fig:displacement}a~and~d for Ecoflex~OO-30 in the undeformed configuration ($\lambda_1 = 1$) at $70$~Hz: the symmetric component of $u_1$ and antisymmetric component of $u_3$ host waves of different wavelengths.
To isolate the different guided modes propagating in the strip, a Singular Value Decomposition (SVD) is applied to the symmetrised monochromatic complex displacement maps. 
For each excitation frequency, the right singular vectors associated with sufficiently large singular values are interpreted as modal displacement fields. 
A spatial Fourier transform along the longitudinal direction $x_1$ is then performed on each of these fields. 
Combining all excitation frequencies and summing the Fourier transforms from the principal singular values yields an intensity map $\mathrm{I}$ in the reciprocal space $\left(\omega, \mathrm{Re}\left(k^{\mathrm{ex}}\right)\right) = \left(\omega, k_r^{\mathrm{ex}}\right)$ displayed in Fig.~\ref{fig:displacement}b, which constitutes the real part of a dispersion diagram.
The imaginary part $k_i^{\mathrm{ex}}(\omega/2\pi)$ visible in Fig.~\ref{fig:displacement}c is extracted by fitting an exponentially decaying envelope to the spatial amplitude of the corresponding modal displacement field ({\it i.e.}, the right singular vector), from which an attenuation length is deduced.

These procedures provide the full complex dispersion relations of the symmetric and antisymmetric in-plane polarized guided waves propagating in the strip.
The first symmetric and antisymmetric in-plane modes are reliably identified across almost the entire frequency range. 
At high frequencies and for higher-order modes however, the intensity map in Fig.~\ref{fig:displacement}b shows information close to $k=0$.
In such cases, the measured displacement corresponds to the spatial average of the field rather than a true propagating wave, resulting in an apparent mode of very long wavelength.

\section{Theoretical framework}
\label{section:theory}
We describe the propagation of incremental guided elastic waves in nearly-incompressible, viscoelastic soft materials under static elongation using a model that takes into account pre-stress, using the acoustoelastic theory \cite{ogden_1997,destrade_2007}, and the material's rheology \cite{delory_2023a,berjamin2025}.
In the following, we briefly recall the key elements of this framework, detailed in references \cite{delory_2023a,delory_2023b}, that are necessary to interpret the experimentally extracted dispersion relations and to relate them to the mechanical properties of the material.

\subsection{In-plane guided elastic waves in nearly incompressible soft materials}
In a homogeneous isotropic elastic medium, such as the silicone elastomers studied here, three bulk wave polarizations exist: one longitudinal and two transverse. 
Introducing the Lamé coefficients $\lambda$ and $\mu$, the longitudinal wave propagates with velocity $V_L = \sqrt{(\lambda + 2\mu)/\rho}$, whereas the two transverse (shear) waves propagate with velocity $V_T = \sqrt{\mu/\rho}$.
For nearly-incompressible materials, that is to say with a Poisson ratio $\nu \to 1/2$, $\lambda \gg \mu$ so that  $V_L \gg V_T$. 
For the elastomers we consider, we indeed find $V_L \approx \qty{1000}{\meter\per\second}$\ and $V_T \approx \qtyrange[range-phrase = -]{1}{10}{\meter\per\second}$ confirming that shear waves dominate the low-frequency dynamics of interest.

We first consider the presence of boundaries $x_2=\pm h/2$, corresponding to a plate geometry, that leads to guided wave propagation.
Shear waves polarized parallel to the plate interfaces do not interact with other polarizations and create the shear horizontal ($S\!H$) spectrum, whereas longitudinal and in-plane transverse waves couple to form the Lamb wave family. 
In the low-frequency regime, only the three fundamental modes are present: the shear horizontal mode ($S\!H_0$), and the first symmetric ($S_0$) and antisymmetric ($A_0$) Lamb modes. 
Interestingly, in this limit, the $S\!H_0$ mode travels at the shear velocity $V_T$, while the $S_0$ mode propagates at the plate velocity $V_P = 2V_T$, reflecting the shear-dominated nature of the material even if the displacement of this mode appears to be longitudinally polarized.

Introducing a finite width along $x_3$ to move from the plate to a strip imposes additional boundary conditions, which couple the $S_0$ and $S\!H_0$ modes. 
This interaction gives rise to a set of in-plane polarized guided modes, observed in our experiments, that are analogous to Lamb waves in a hypothetical medium where $V_L = 2 V_T$~\cite{laurent_2020}. In this analogy with the Lamb modes, they decompose onto a family of antisymmetric modes (denoted $A^\prime_n$) and symmetric ones ($S^\prime_n$).
We obtain the theoretical dispersion relation of this set of modes using a semi-analytical approach based on the spectral collocation method~\cite{trefethen_spectral_2000,weideman_matlab_2000}.
We introduce the displacement vector $\boldsymbol{u} = \boldsymbol{u}\left(k,x_2,x_3,\omega\right)e^{i(k x_1 - \omega t)}$ and look for solutions of the wave equation
\begin{equation}
C_{jikl} \frac{\partial^2 u_k}{\partial x_j \partial x_l} + \rho \omega^2 u_i = 0,
\label{eq:wave}
\end{equation}
where $\boldsymbol{C}$ is the fourth order elasticity tensor, subjected to zero traction boundary conditions at the strip edges.
After discretization along the strip cross-section, the problem reduces to an algebraic system relating the displacement field $\boldsymbol{u}$ to the longitudinal wavenumber $k$ and angular frequency $\omega$.
This system can be solved either for the eigenpair $(\omega^2, \boldsymbol{u})$ parametrized by $k$, or for $(k,  \boldsymbol{u})$ parametrized by $\omega$. 
The latter approach is particularly suited to our case as we are interested in determining the complex-valued wavenumber $k$ for real frequencies.
Our implementation is freely available~\cite{kiefer_2024_10641373}, and has been used to study guided waves in plates and strips~\cite{kiefer_elastodynamic_2022,delory_2023b}.
\begin{figure}[t]
	\centering
	\includegraphics[width=\linewidth]{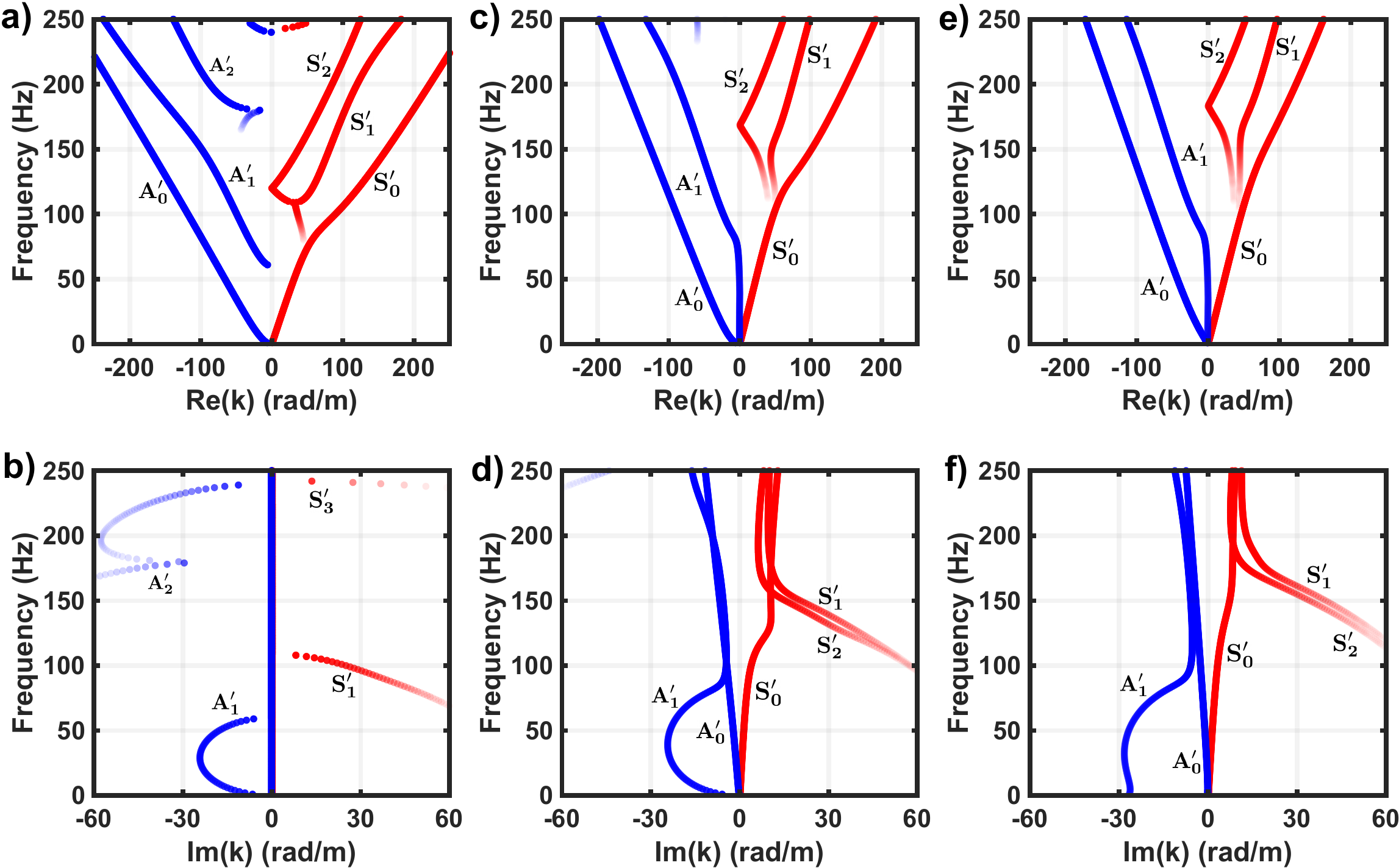}
	\caption{\textbf{In-plane guided elastic modes in a nearly-incompressible soft strip ---}  (a,b) Theoretical dispersion curves of a linear elastic strip with thickness $h=1$\,mm, width $b=5$\,cm, density $\rho = 1$\,g/cm$^3$, transverse velocity $V_T = 6$\,m/s, and longitudinal velocity $V_L = 1000$\,m/s. The real and imaginary parts of the wavenumbers correspond to the propagative and lossy components, respectively. The first in-plane symmetric modes are shown in red, and the first in-plane antisymmetric modes are shown in blue with negative wavenumbers. The evanescent behavior of the guided modes is encoded in their transparency.
		(c,d) The same curves with the viscoelastic behavior of Eq.~\ref{eq:fractionalVoigt} included ($\tau = 1$\,ms, $n=0.15$).  
		(e,f) Curves with both viscoelasticity and a uniaxial elongation of $\lambda_1 = 1.3$ ($\alpha = 0.3$, $\beta^\prime=0$).
		\label{fig:SCM}
	}
\end{figure}
It yields the full set of symmetric and antisymmetric in-plane polarized guided modes, which we represent in figure \ref{fig:SCM}(a-b) for a strip with similar mechanical properties ($\lambda$, $\mu$ and $\rho$) and dimensions as that used in our experiments. The symmetric modes are represented in red with a positive wavenumber, and the antisymmetric ones in blue with a negative wavenumber (and similarly for the imaginary part). The transparency in the real part of $k$ reflects the magnitude of the imaginary part, so as to focus attention on the least attenuated modes, notably the fundamental modes $S^\prime_0$, whose displacement is mainly along $x_1$ (see figure~\ref{fig:displacement}a),
and $A^\prime_0$, whose displacement is mainly along $x_3$ (see figure~\ref{fig:displacement}d).
Higher-order modes only appear beyond their respective cut-off frequencies, below which they exist as evanescent branches shown in panel (b).
This set of guided modes acts as multiple observables that should help us in retrieving the properties of the elastomer under study.

While a strip acts as a dispersive propagating medium, this behavior arises primarily from the geometrical guiding. The next section focuses on how the frequency-dependent mechanical properties of the material itself can be quantified using the extracted dispersion relations.

\subsection{Frequency-dependent properties of soft elastomers: viscoelasticy}
The shear modulus of a viscoelastic material is generally complex, $\mu = \mu^\prime + i \mu^{\prime\prime}$, with a real part $\mu^\prime$ representing the elastic response and an imaginary part $\mu^{\prime\prime}$ representing viscous dissipation. 
Causality and the Kramers-Kronig relations~\cite{lee1990experimental} imply that the presence of dissipation necessarily introduces a frequency dependence of $\mu$, and conversely, any frequency dependence of the shear modulus is accompanied by an imaginary component. 
In other words, the viscoelastic nature of the material gives rise to frequency-dependent mechanical parameters, in particular the shear modulus $\mu$.
Not all commonly used rheological models satisfy the constraints imposed by causality: in particular, the Kramers-Kronig relations require a specific connection between the real and imaginary parts of the shear modulus. 
Independent rheological measurements on soft elastomers further indicate that the frequency dependence of $\mu$ often follows a power-law with a non-integer exponent. 
To capture both the frequency-dependent elasticity and dissipation consistently, we therefore adopt a fractional-derivative model. 
We choose the fractional Kelvin-Voigt model~\cite{smit1970rheological,bagley1983fractional,liu_2014,rolley2019flexible,mainardi2022fractional,sharma_2023,parker_2019}, in which the complex shear modulus reads
\begin{equation}
\mu(\omega) = \mu_0 \left[ 1 + (i \omega \tau)^n \right],
\label{eq:fractionalVoigt}
\end{equation}
where $\mu_0$ is the static shear modulus, $\tau$ is a characteristic relaxation time, and $n$ is the fractional exponent. 
The storage modulus (real part) and loss modulus (imaginary part) of $\mu$ therefore vary with the angular frequency $\omega$ and are fully determined by these three parameters. 

We illustrate the influence of considering a frequency-dependent shear modulus by computing the dispersion relation of the strip's in-plane modes using our semi-analytical method where the elasticity tensor now depends on $\omega$.
We choose the same parameters as in figure \ref{fig:SCM}(a-b) but consider the viscoelastic parameters $\tau = \qty{1}{\milli\second}$ and $n = 0.15$.
Taking the material's viscoelasticity into account has several visible consequences (figure~\ref{fig:SCM}c-d). First, the dispersion branches steepen compared to the purely elastic case, consistent with a shear modulus that increases with frequency; this stiffening also shifts the cut-off frequencies of higher-order modes toward higher frequencies. Second, all modes now exhibit a non-zero imaginary part of $k$, reflecting the energy dissipation introduced by viscoelasticity — including modes that were purely propagative in the elastic case, whose attenuation branches are now clearly visible in panel (d). Third, the near-degeneracy previously observed between $S^\prime_1$ and $S^\prime_2$ is resolved into a clear separation between the two branches, and the zero-group-velocity (ZGV) point disappears~\cite{laurent_2020,lanoy_2020,delory_2022}: viscoelasticity lifts the degeneracy between the two branches. Consistently, the formerly degenerate imaginary parts split into two distinct and distinguishable curves. A similar behaviour is visible for branches $A^\prime_1$ and $A^\prime_2$.

\subsection{Hyperelastic and acoustoelastic behavior under uniaxial elongation}
In addition to their frequency-dependent response, the elastodynamics of soft elastomers are strongly influenced by prestress.
We now describe how an underlying static deformation influences the propagation of elastic waves using the incremental approach of Ogden and Destrade~\cite{ogden_1997,destrade_2007}.
The mechanical behaviour of nearly incompressible isotropic materials is described by a strain-energy density $W$ which depends on the first three invariants of the left (right) Cauchy-Green tensor
\begin{eqnarray}
I_1 &=& \lambda_1^2 + \lambda_2^2 + \lambda_3^2,\\
I_2 &=& \lambda_1^2 \lambda_2^2 + \lambda_1^2 \lambda_3^2 + \lambda_2^2 \lambda_3^2,\\
I_3 &=& \lambda_1^2 \lambda_2^2 \lambda_3^2,
\end{eqnarray}
where $(\lambda_1, \lambda_2, \lambda_3)$ are the principal stretch ratios.
As previously reported for Ecoflex OO-30, the hyperelastic behavior of the strip in uniaxial extension is well captured by the compressible Mooney-Rivlin constitutive law~\cite{delory_2023a,delory_2023b} which introduces two parameters: the static shear modulus $\mu_0$ and $\alpha$ that both appear in the expression of $W$ (see appendix~\ref{section:calculus}).
Assuming incompressibility, that is $I_3 = 1$, these parameters can be extracted from the stress-strain curve recorded during a simple tension test.
For this deformation, the stretch ratios are $(\lambda_1, \lambda_1^{-1/2}, \lambda_1^{-1/2})$ and the engineering stress $\sigma$ is given by
\begin{equation}
\sigma = \frac{\partial W}{\partial \lambda_1} = 2 \left(\lambda_1 - \frac{1}{\lambda_1^2}\right) \frac{\mu_0}{2} \left( 1 - \alpha + \frac{\alpha}{\lambda_1} \right).
\label{eq:sigma}
\end{equation}
Following the methodical procedure detailed in reference~\cite{destrade2017methodical}, we represent the data in the Mooney space: we plot the corrected stress function $g = \sigma / \left(2\lambda_1 - 2/\lambda_1^2\right)$ as a function of $1 / \lambda_1$. In this space, $g$ depends linearly on $1/\lambda_1$ with an offset.
\begin{figure}[t]
	\centering
	\includegraphics[width=\linewidth]{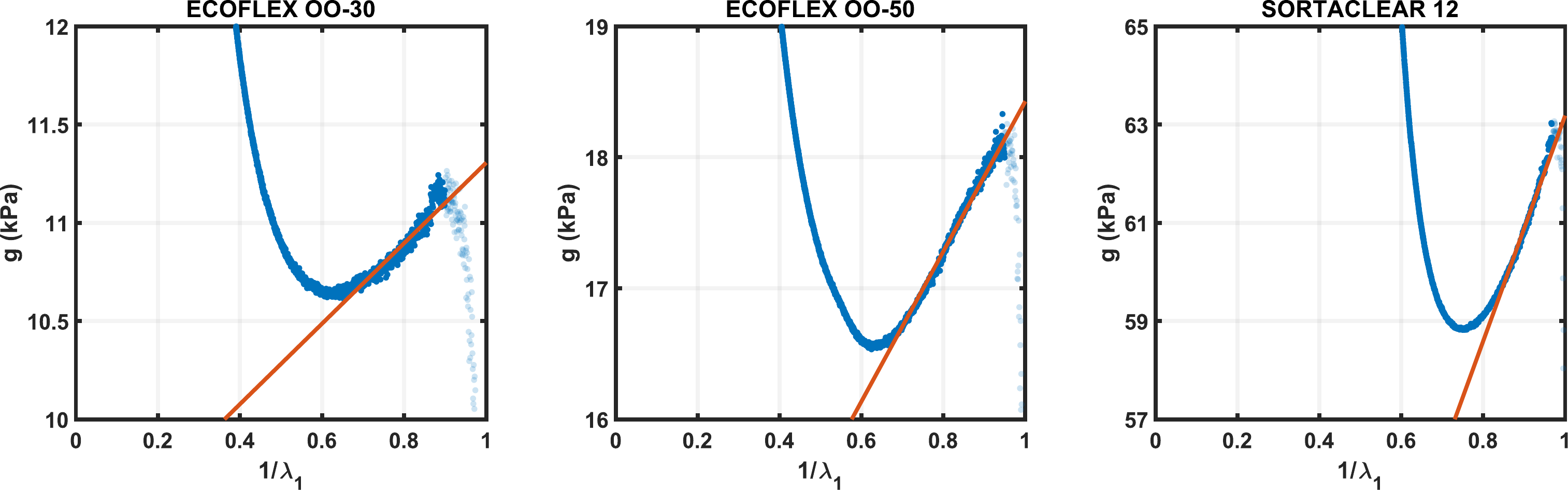}
	\caption{\textbf{Results of the traction test in the Mooney space ---} The corrected stress applied to the strip $g = \sigma/(2\lambda_1 - 2\lambda_1^{-2})$ is plotted as a function of $1/\lambda_{1}$. The parameters $\mu_0$ and $\alpha$ are fitted on the linear section that corresponds to the domain of validity of the Mooney-Rivlin model. Datapoints for $\lambda_1\approx1$ are discarded due to the strip not being under enough tension.  }
	\label{fig:mooneyplot}
\end{figure}
Figure~\ref{fig:mooneyplot} shows the Mooney plots extracted from uniaxial tensile tests performed on the three elastomers.
As evidenced by equation \eqref{eq:sigma}, the linear portion of the curve indicates the range of validity of the Mooney-Rivlin model that corresponds to the small to moderate strain regime. Additionally, measurement errors in the tensile tests prevent us from using the data for $\lambda_1 \approx 1$.
For Ecoflex~OO-30 and OO-50 the model is appropriate when $\lambda_1 \in [1.1,1.4]$, while for SortaClear~12 we obtain $\lambda_1 \in [1.05, 1.25]$.
Linear least-square fits within the validity range of each elastomer yield the parameters $\mu_0$ and $\alpha$, summarized in Table~\ref{tab:parameters}.

Having extracted the hyperelastic material parameters from static tests, we discuss the propagation of small amplitude waves on top of an underlying static uniaxial deformation.
Such incremental waves also obey a wave equation, namely:
\begin{equation}
    C_{jikl}^\omega\left(\lambda_1,\mu_0,\alpha,\omega,\tau,n,\beta^\prime\right) \frac{\partial^2 u^\prime_k}{\partial x_j \partial x_l} + \rho \omega^2 u^\prime_i = 0,
\label{eq:waveincr}
\end{equation}
where $\boldsymbol{u}^\prime$ stands for the incremental displacement, and the elasticity tensor has been replaced by the visco-acoustoelastic tensor $\boldsymbol{C}^\omega$.
This fourth-order tensor combines the push-forward of $\boldsymbol{C} $ in the deformed configuration, $\boldsymbol{C}^0$, and a frequency-dependent contribution, stemming from the fractional Kelvin-Voigt model, as detailed in Appendix~\ref{section:calculus} \cite{delory_2023a,berjamin2025}.
Note that the coupling between acoustoelasticity and viscoelasticity leads to the appearance of an additional fit parameter $\beta^\prime$.
Equation \eqref{eq:waveincr} can be solved using our semi-analytical method once we determine $\boldsymbol{C}^\omega$ as detailed in reference \cite{delory_2023b}.
We illustrate the influence of prestress by computing the dispersion relations of a strip, taking into account a typical example of hyperelastic and viscoelastic parameters ($\alpha = 0.3$, $\beta^\prime = 0$), subjected to an elongation $\lambda_1 = 1.3$ (figure \ref{fig:SCM}e-f).
Adding a pre-strain to the ribbon introduces further modifications to the dispersion relations (figure~\ref{fig:SCM}e-f). The most naive expectation — that stretching uniformly stiffens all modes — is only partially verified. While $A^\prime_0$, $S^\prime_1$ and $S^\prime_2$ are noticeably shifted toward higher wavenumbers, consistent with an increased apparent rigidity, the modes $S^\prime_0$ and $A^\prime_1$ appear comparatively unaffected. This selective sensitivity to pre-strain reflects the fact that different modes sample the elastic properties of the strip differently, depending on their polarization and spatial distribution of strain energy. Additionally, the imaginary parts of $k$ tend to decrease slightly under pre-strain, suggesting a modest reduction in attenuation, though this effect remains subtle and not uniformly distributed across modes.

We have evidenced that waves propagating in our pre-stressed strip are strongly influenced by the guiding geometry, the frequency-dependent viscoelasticity, and the elongation-dependent acoustoelasticity.
We now translate this sensitivity and the theoretical framework that takes into account these three contributions, into a minimization procedure for mechanical characterization.

\section{Mechanical characterization}
We  describe how to determine an elastomer's mechanical properties by matching the measured dispersion relations of guided waves in a strip, for various elongations $\lambda_1$, to predictions from the visco-acoustoelastic incremental theory described in the previous section.
The combination of viscoelasticity and acoustoelasticity allows us to exploit dispersion relations in both undeformed and statically pre-stressed configurations of the strip, providing more independent observables than could be obtained from the undeformed state alone.

The characterization procedure consists in computing the theoretical complex-valued wavenumber at each pulsation from the wave equation~(\ref{eq:waveincr}), under the strip's boundary conditions, and comparing it with the experimentally measured wavenumber $k^\mathrm{ex}(\omega)$. 
This comparison involves five material parameters: the static shear modulus $\mu_0$, the Mooney-Rivlin model coefficient $\alpha$, the viscoelastic parameters $\tau$ and $n$, and the term coupling the viscosity and the acoustoelasticity $\beta^\prime$.
In this contribution, we focus on determining the viscoelastic parameters.
We obtain the values of $\mu_0$ and $\alpha$ by fitting the data from uniaxial extension tests in the Mooney-space (figure \ref{fig:mooneyplot}), and perform a minimization procedure to determine the viscoelastic parameters $\tau$, $n$, and the coupling parameter $\beta^\prime$.

\subsection{Parameter estimation\label{section:estimation}}

Estimating the set of parameters $\boldsymbol{\theta} = [\tau, n, \beta^\prime]$, requires to quantify the agreement between the experimental and theoretical dispersion relations.
To do so, we introduce an error function $\mathcal{E}\left(\boldsymbol{\theta}\right)$ that is the sum of the contributions associated to the error on the real and imaginary parts of the wavenumber.
\begin{figure}[t]
	\centering
	\includegraphics[width=\linewidth]{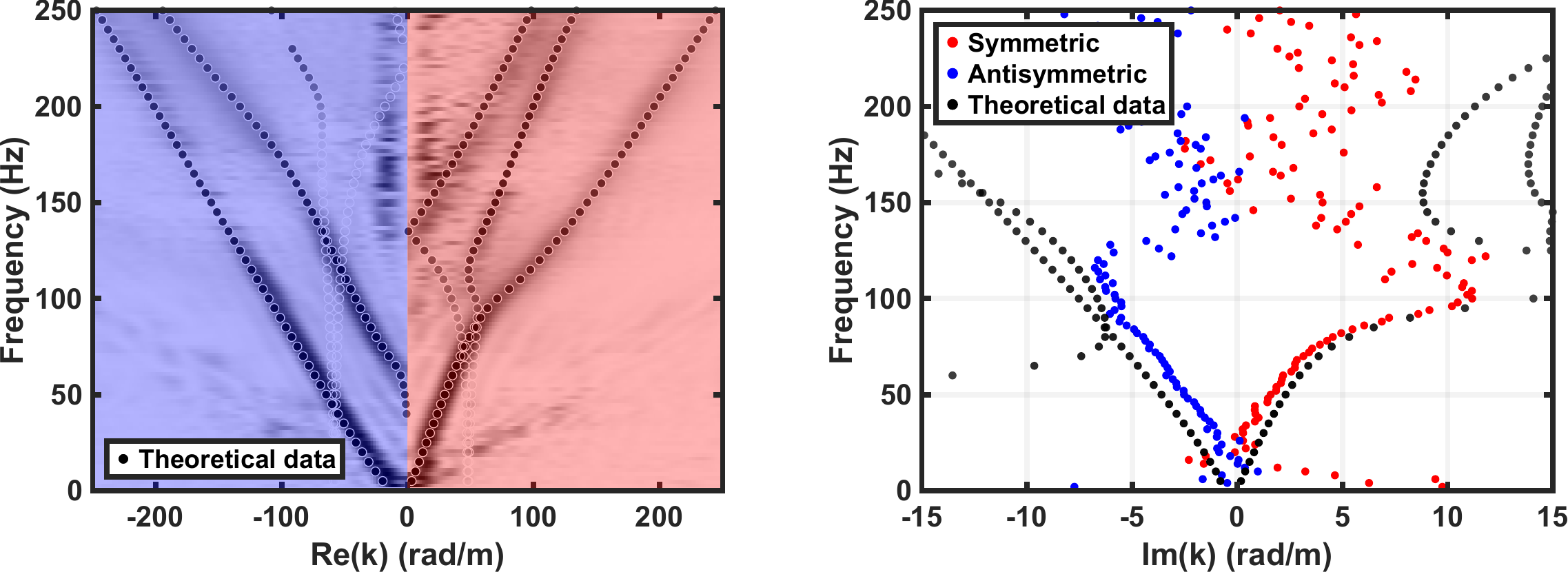}
	\caption{\textbf{Comparison between the experimental dispersion diagrams and the theoretical dispersion curves of Ecoflex OO-30 in the undeformed configuration ---} a) The likelihood of the theoretical estimation is calculated by mapping the theoretical dispersion curves on the experimental intensity maps (constituting the real part of the dispersion). Here the likelihood is close to maximal and corresponds to a minimal error function. The evanescent behavior of the guided waves is encoded in the transparency of the datapoints. b) The likelihood of the imaginary datapoints is calculated using a gaussian error function.}
	\label{fig:interpolation}
\end{figure}
The error on $\textrm{Re}(k)$ is computed by mapping the theoretical dispersion curves onto the experimental intensity maps $\textrm{I}^\textrm{ex}[f, \textrm{Re}(k)]$, as illustrated in Fig.~\ref{fig:interpolation}:
\begin{equation}
\mathcal{E}_r(\boldsymbol{\theta}) = 
-\sum_{\lambda_1,f} 
\textrm{I}^\textrm{ex}\left[f, \textrm{Re}\left(k^\mathrm{th}(\lambda_1,f,{\boldsymbol\theta})\right)\right],
\label{eq:realerror}
\end{equation}
where we sum on all frequencies $f$ probed in our experiments, and on all elongations $\lambda_1$ within the small to moderate strain range. This method avoids mode matching and is appropriate for noisy measurements by overlooking outliers.

The second contribution accounts for the imaginary part $\textrm{Im}(k)$.
To reduce the influence of outliers due to noisy data, we compute a Gaussian-weighted distance between the theoretical and experimental values of the principal mode of each symmetry:
\begin{equation}
\mathcal{E}_i(\boldsymbol{\theta}) = 
-\sum_{\lambda_1,f} 
\exp\Bigg[-\frac{\left(k_i^{\mathrm{ex}}(\lambda_1,f) - \textrm{Im}\left(k^\mathrm{th}(\lambda_1,f,{\boldsymbol\theta})\right)\right)^2}{2\sigma^2}\Bigg],
\label{imaginaryerror}
\end{equation}
with $\sigma \approx 2$\,rad/m: this value is chosen to be consistent with the real part of the experimental dispersion diagram such as the one in Fig.~\ref{fig:displacement}b, where it corresponds to the approximate width of the dispersion curves. It also disregards the outliers such as the high-frequency ones in Fig.~\ref{fig:displacement}c.

The total error function is finally defined as
\begin{equation}
\mathcal{E}(\boldsymbol{\theta}) = \mathcal{E}_r(\boldsymbol{\theta}) + w_\mathrm{i}\,\mathcal{E}_i(\boldsymbol{\theta}),
\label{eq:error}
\end{equation}
where a reduced weight $w_\mathrm{i} \approx 0.3$ is applied to the imaginary part, empirically chosen to account for its higher noise level (see appendix \ref{section:additional}).

Minimizing $\mathcal{E}(\boldsymbol{\theta})$ yields the optimal set of parameters describing the viscoelastic and acoustoelastic response of the elastomer.
To avoid local minima in this noisy parameter landscape, a stochastic algorithm is preferred. 
Here, we employ Matlab's \texttt{genetic algorithm}. 
To quantify the robustness of the parameter estimation, the characterization procedure was repeated multiple times.
\begin{table}[t]
	\centering
	\setlength{\tabcolsep}{11pt}
	\renewcommand{\arraystretch}{1.3}
	\begin{tabular}{|l|c|c|c|c|c|}
		\hline
		\multirow{2}{*}{ } & \multicolumn{2}{c|}{Traction test} & \multicolumn{3}{c|}{Characterization procedure} \\
		\cline{2-6}
		& $\mu_0$\,(kPa) & $\alpha$ & $n$ & $\tau$\,(ms) & $\beta^\prime$ \\ 
		\hline\hline
		ECOFLEX OO-30 & 22.6 & 0.18 & 0.20~$\pm$~0.04 & 2.9~$\pm$~1.3 & 0.70~$\pm$~0.33\\
		\hline
		ECOFLEX OO-50 & 37.0 & 0.32 & 0.24~$\pm$~0.02 & 2.8~$\pm$~0.6 & 0.67~$\pm$~0.08 \\
		\hline
		SORTACLEAR 12 & 127 & 0.37 & 0.08~$\pm$~0.02 & 0.06~$\pm$~0.04 & 0.37~$\pm$~0.29 \\ 
		\hline
	\end{tabular}
	\caption{Mechanical parameters obtained from traction tests and characterization procedure for the three example elastomers.}
	\label{tab:parameters}
\end{table}
Due to the roughness of the parameter landscape, slightly different sets of parameters are obtained at each iteration. The mean values of the optimized parameters for each elastomer are reported in Table~\ref{tab:parameters}, providing a reliable characterization of their viscoelastic and acoustoelastic properties. The robustness of the algorithm concerning the parameter landscape boundaries allows for an almost blind estimation of the parameters.

\subsection{Results of the characterization procedure}
\begin{figure}[t]
	\centering
	\includegraphics[width=\linewidth]{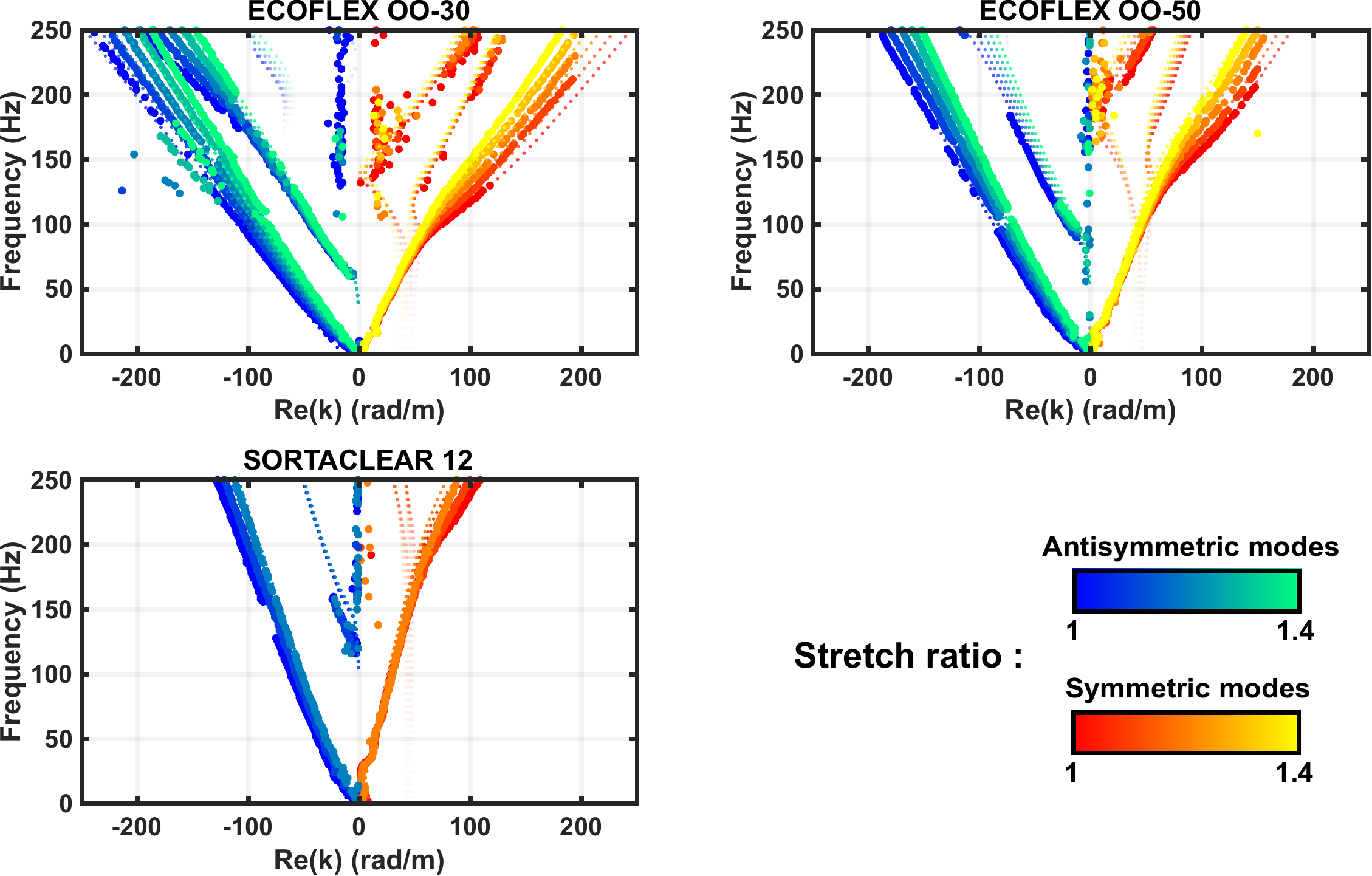}
	\caption{
		\textbf{Dispersion diagrams for all materials ---} Experimental curves are shown as thick dotted lines, while theoretical curves are represented with thin dotted lines. Symmetric modes are plotted with positive wavenumbers and antisymmetric modes with negative ones. Corresponding stretch ratios are indicated by the colorbar. The stretch ratio is restricted to the validity of the Mooney-Rivlin model: $\lambda_1 \in [1, 1.4]$ for Ecoflex and $\lambda_1 \in [1, 1.25]$ for Sortaclear.
	}
	\label{fig:dispersion}
\end{figure}
For each elastomer sample, the resulting optimized theoretical dispersion curves are directly compared to the experimental ones in Fig.~\ref{fig:dispersion}.
The real part of the experimental longitudinal wavenumbers $k_r^{\mathrm{ex}}(\omega/2\pi)$ is extracted from the data displayed in Fig.~\ref{fig:displacement} as the locations of the maxima in the intensity map.
The imaginary counterpart is studied in appendix~\ref{section:additional}.
The comparison with the numerical predictions demonstrates good agreement for all three elastomers considered, highlighting several key aspects of the method. 
First, although the experimental measurements do not capture all guided modes at all frequencies due to noise or weak signals, the theoretical predictions fill in these gaps convincingly, and the few experimentally observable points align well with the corresponding branches of the computed dispersion curves. 
Second, the frequency-dependence of the shear modulus due to viscoelasticity, which produces noticeable changes in the dispersion diagrams as the excitation frequency varies, is well captured by the theoretical curves that accurately follow the experimental points. 
Third, the application of uniaxial pre-stress does not affect all modes in the same manner. 
The numerical framework correctly predicts mode-dependent shifts in the dispersion relations with increasing stretch ratio, which is consistent with the observations in the experiments. 

Taken together, these observations confirm that the combination of viscoelasticity and acoustoelasticity in the model provides a faithful representation of the complex behavior of the guided waves in pre-stressed soft elastomer strips. 
Inversely, measuring the dispersion diagram of modes propagating in an elastic strip under stress provides enough information to retrieve viscoelastic material parameters that allow to accurately describe the experimental data.

\subsection{Comparison to conventional rheometry}
\begin{figure}[t]
    \centering
    \includegraphics[width=\linewidth]{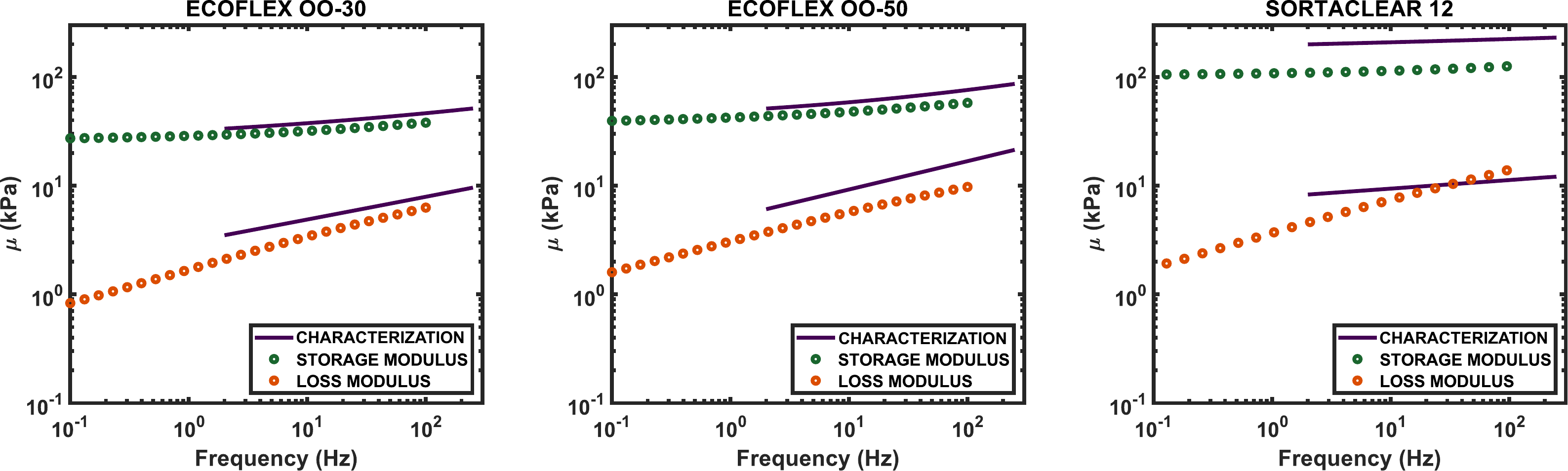}
    \caption{\textbf{Comparison of guided wave identification with conventional rheometry ---} Measurements of the shear modulus (storage part and loss part) on an Anton-Paar MCR501 rheometer in the plate-plate configuration are compared to the output of the characterization procedure.}
    \label{fig:rheology}
\end{figure}

Finally, we compare the optimal viscoelastic parameters, given in table \ref{tab:parameters}, to that obtained through conventional rheological measurements performed on an Anton-Paar MCR501 rheometer in plate-plate configuration (circles in figure~\ref{fig:rheology}).
We calculate the storage and loss moduli from the estimated parameters using equation \eqref{eq:fractionalVoigt} and represent them as lines in figure~\ref{fig:rheology}. 
As the two experimental approaches probe different frequency ranges, the comparison focuses on the overlapping domain, {\it i.e.} from \qty{1}{\hertz} to \qty{100}{\hertz}.
For Ecoflex~OO-30 and Ecoflex~OO-50, the storage and loss moduli optimized from the guided-wave method are in good overall agreement with the rheometer measurements over their respective frequency ranges. In particular, the slope of the loss modulus, governed by the fractional exponent $n$, is consistently recovered by both methods. The relaxation time $\tau$, which controls the frequency at which the storage modulus begins to plateau, shows larger discrepancies between the two methods — up to one order of magnitude depending on the sample. The relaxation time $\tau$, which controls the frequency at which the storage modulus begins to bend upwards, shows larger discrepancies between the two methods — up to one order of magnitude depending on the sample. While the origin of this disagreement is not fully understood, both methods nevertheless capture a similar inflection in the storage modulus, suggesting that the characteristic timescale is at least qualitatively recovered. The remaining deviations, quantified in Table~\ref{tab:parameters}, may also originate from slight inhomogeneities in the static elongation of the strips under their own weight, which can locally modify the effective mechanical properties (see appendix~\ref{section:weight} for a detailed discussion).

For Sortaclear~12, the agreement is less satisfying. 
The larger relative standard deviations of the optimized parameters (table~\ref{tab:parameters}) reflect both experimental variability and material-specific effects. 
In particular, the shorter pot life of Sortaclear~12 compared to the Ecoflex series may have induced subtle differences in polymerization between the sample used for rheometry (solidified between two plates at the millimeter scale) and the strip used in our guided-wave experiments (solidified in a free-surface mold at centimeter scale). 
Such differences can result in residual stresses or local variations in microstructure, affecting the measured viscoelastic responses.

Despite these discrepancies, the guided-wave method provides a reliable estimation of the rheological parameters across a broad frequency range, in a single experimental configuration. 
Importantly, it simultaneously accounts for the effect of pre-stress on the mechanical response, which is not captured in standard rheometry. 
The method is highly versatile: the frequency range can be extended using stroboscopic techniques or high-speed imaging. Nevertheless, at high frequencies, the upper limit is set by wave attenuation: as the waves become increasingly damped, near-field effects close to the source and propagating attenuated waves become difficult to disentangle, introducing errors in the extracted parameters. The numerical framework also allows for the substitution of the fractional-derivative and Mooney-Rivlin models with alternative rheological or hyperelastic descriptions, enabling the characterization of a wide variety of soft materials.
Overall, this comparison demonstrates that guided elastic waves in pre-stressed strips constitute a powerful and flexible approach for broadband rheology of elastomers, complementing the conventional frequency range of rheometric techniques.

\section{Conclusion}
We demonstrate that guided elastic waves provide a powerful alternative to conventional mechanical tests for characterizing soft elastomers. 
Traditional approaches rely on simultaneous measurements of force and displacement, whereas wave propagation inherently encodes the mechanical response of the material, offering a ``free'' dynamic probe. 
Because waves are intrinsically dynamic, they naturally capture the frequency-dependent behavior of the material and, by virtue of Kramers-Kronig relations, also provide access to the viscous effects.  

By shaping the material as a thin strip, the guided geometry introduces multiple propagative modes, thereby increasing the number of independent experimental observables and improving the reliability of the characterization. 
The experimental setup is extremely simple and yet sufficiently versatile to probe both the frequency-dependent viscoelasticity and the elongation-dependent acoustoelasticity of the material, providing a more complete picture of its mechanical properties.  

We have applied this approach to several hyperelastic silicone elastomers from the Smooth-On range. For all tested materials, the extracted mechanical parameters are consistent with independent rheological measurements, and the method has allowed extension of the frequency range by increasing the upper bound compared to conventional rheometry.

Several avenues remain for improvement. Currently, parameter estimation relies on a numerical minimization procedure to find the set of material parameters that best reproduces the experimental dispersion curves. Developing a rigorous inverse method capable of directly extracting the frequency-dependent shear modulus $\mu(\omega)$ from the data would streamline this process. Additionally, implementing mechanical excitation directly on the traction test itself would enable hyperelastic and viscoelastic measurements on the same sample, further reducing experimental variability.
Overall, this work paves the way for robust, versatile, and potentially low-cost approaches to broadband rheology of soft materials. Extensions to real-time parameter estimation and clinical applications, such as elastography of muscles or tendons that fully account for waveguiding effects, appear within reach~\cite{sigrist_2017, li_2017, bilston_2018, crutison_2022,croquette2025exploring}.

\bibliographystyle{unsrt}
\bibliography{biblio}

\appendix

\section{Hyperelasticity and acoustoelasticity}
\label{section:calculus}

We describe the hyperelastic behavior of the strip with a compressible Mooney-Rivlin model. Introducing the bulk modulus $\kappa = \lambda + 2\mu/3$\,, the stretch ratios along all directions $\left(\lambda_1, \lambda_2, \lambda_3\right)$ as well as the invariants $I_1 = \lambda_1^2 + \lambda_2^2 + \lambda_3^2\,,$ $I_2 = \lambda_2^2\lambda_3^2 + \lambda_1^2\lambda_3^2 + \lambda_1^2\lambda_2^2$\,, and $I_3 = J^2 \lambda_1^2\lambda_2^2\lambda_3^2$\,, the energy function $W$ of the Mooney-Rivlin model writes 
\begin{equation}
        W = \frac{\mu_0}{2}\Bigg[(1-\alpha)\left(\frac{I_1}{I_3^{1/3}}-3\right)+\alpha\left(\frac{I_2}{I_3^{2/3}}-3\right)\Bigg] \\ + \frac{\kappa}{2}(\sqrt{I_3}-1)^2.
    \label{eq:StrainEnergy}
\end{equation}
The parameter $\alpha$ encodes the deviation of the material's response from the Gaussian chain statistical theory: a Neo-Hookean solid would correspond to $\alpha=0$~\footnote{This equation is equivalent to its formulation in terms of the Landau coefficients $A$ and $D$~\cite{destrade2010third,destrade2017methodical}.}\,.

Then, the infinitesimal elastodynamics of this elongated viscoelastic solid can be fully described with a new elastic tensor $\boldsymbol{C}^0$, following the incremental approach by Ogden and Destrade~\cite{ogden_1997,destrade_2007}. As the static deformation breaks some of the symmetries of this tensor, it differs from the conventional elasticity tensor used for isotropic or even transverse isotropic solids. As explained in~\cite{delory_2023a}, the sum of this tensor accounting for acoustoelasticity with a viscoelastic term from equation~(\ref{eq:fractionalVoigt}) gives the components of the resulting tensor $\boldsymbol{C}^\omega$ that appears in the wave equation~\eqref{eq:waveincr}

\begin{equation}
    C^\omega_{ijkl} = C^0_{ijkl} + \mu_0\left(i\omega\tau\right)^n \left(1 +\beta^\prime\,\frac{\lambda_i^2+\lambda_j^2-2}{2}\right)I_{ijkl}
    \label{eq:Comega}
\end{equation}

\noindent where $I_{ijkl}=(\delta_{ik}\delta_{jl}+\delta_{il}\delta_{jk})$ is a function of the Kronecker delta $\delta_{ik}$.
The $C^0_{ijkl}$ components are then derived from equation~(\ref{eq:StrainEnergy}) using the formulas:

\begin{eqnarray}
    C^0_{iijj} &= \frac{\lambda_i\lambda_j}{J} \, W_{ij}&\quad \\
    C^0_{ijji} &= \frac{\lambda_i^2}{J} \, \frac{\lambda_i W_i - \lambda_j W_j}{\lambda_i^2-\lambda_j^2} & (i\neq j, \lambda_i\neq\lambda_j)\\
    C^0_{ijji} &= \frac{C^0_{iiii}-C^0_{iijj}+\lambda_i W_i/J}{2} & (i\neq j, \lambda_i=\lambda_j)\\
    C^0_{ijij} &= \frac{\lambda_i\lambda_j}{J} \, \frac{\lambda_j W_i - \lambda_i W_j}{\lambda_i^2-\lambda_j^2} & (i\neq j, \lambda_i\neq\lambda_j)\\
    C^0_{ijij} &= \frac{C^0_{iiii}-C^0_{iijj}-\lambda_i W_i/J}{2} & (i\neq j, \lambda_i=\lambda_j)
\end{eqnarray}

\noindent where $\displaystyle W_i = \frac{\partial W}{\partial \lambda_i}$ and $\displaystyle W_{ij} = \frac{\partial^2 W}{\partial \lambda_i \partial \lambda_j}$\,. Additional invariants would be considered for anisotropic materials but the underlying reasoning remains the same~\cite{balzani_2006,peyraut_2010,li_cao_2017,mukherjee_2022}.

\section{Accounting for weight-induced elongation of the sample\label{section:weight}}
The experimental setup presented in this article relies on a vertical strip sample (see Fig.~\ref{fig:setup}).
Even if the strip is not subjected to any imposed stretching, a residual elongation coming from its own weight is present.
In what we would label the undeformed configuration (\emph{i.e.}, with no additional pre-stress induced by the motor), the strip displays a gradient of elongation.
The bottom of the strip is stress free, while the top, which carries the whole weight of the sample, exhibits a maximal elongation.
For Ecoflex~OO-30, the longitudinal stretch ratio varies from $\lambda_1 = 1$, at the bottom of the strip, to $\lambda_1\approx1.04$, at the top. 
For stiffer materials like SORTACLEAR~12, the gravity induced deformation is less of an issue with a maximal elongation of  $\lambda_1\approx1.02$. 
Since the attenuation distance of monochromatic waves depends on the frequency, a low-frequency wave travels further than a high-frequency one: the space-dependent stretch ratio translates into a frequency-dependent one.
In our measurements, we take the average $\lambda_1$ for each imposed elongation as the input for the calculation of the theoretical dispersion curves. 
This approximation is one of the factors that could explain the gap with the conventional rheological measurements in Fig.~\ref{fig:rheology}. 
The influence of the approximation is less significant for elongated configurations as the strip becomes stiffer.

\section{Additional input for parameter estimation\label{section:additional}}
\subsection{Weighting of the imaginary part of the error function}
\begin{figure}[t]
    \centering
    \includegraphics[width=\linewidth]{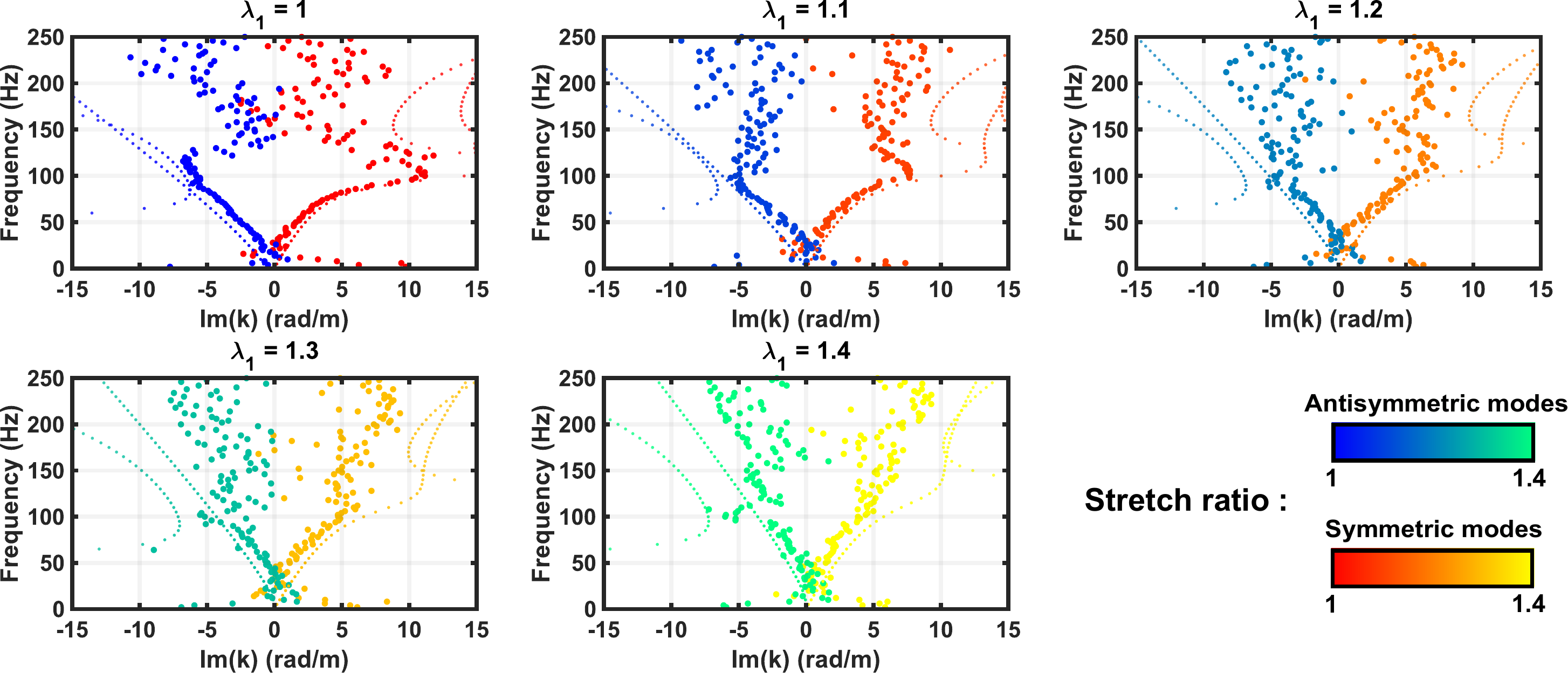}
    \caption{\textbf{Imaginary part of the dispersion curves of Ecoflex OO-30---} For elongations in the domain validity of the Mooney-Rivlin model $\lambda_1 \in [1, 1.4]$, the experimental curves are represented in big dotted lines and the theoretical ones in small dotted lines. Symmetric modes are plotted with positive wavenumbers and antisymmetric modes with negative ones to enhance clarity. Corresponding elongation ratios and symmetries are given by the colorbar.
    }
    \label{fig:imaginary}
\end{figure}

Figure~\ref{fig:imaginary} displays the experimental and optimized imaginary part of the dispersion curves for Ecoflex OO-30 when varying the imposed elongation $\lambda_1$. 
While we expect the imaginary part of the experimental wavenumber to carry more information on the loss modulus of the material than when considering the real part alone, the scatter on the imaginary part is much larger as can be observed by comparing figures~\ref{fig:dispersion} and~\ref{fig:imaginary}.
This lead us to give more weight  to the error function calculated from the real part than to the one calculated from the imaginary part when optimizing the problem, as explained in section~\ref{section:estimation}.

\subsection{Low-frequency wavenumber extraction}
When the wavelength is of the order of the size of the sample, performing a Fourier transform tends to overestimate the real part of the wavenumber.
This behaviour can be observed in the data of figure~\ref{fig:interpolation} for frequencies around $\qty{1}{\hertz}$.
While $\mathrm{Re}(k)$ should tend to $0$ as the frequency decreases according to the theoretical dispersion curves, the experimental wavenumber remains at a non-zero value.

The real part of the wavenumber can be estimated differently in this low-frequency regime.
As there is only one in-plane symmetric mode and one in-plane antisymmetric mode, an exponentially decreasing oscillating envelope can be fitted on the raw symmetrized displacement maps, before Singular Value Decomposition (see section~\ref{section:postprocessing}), instead of the final Fourier transform.
\end{document}